\documentclass[journal]{IEEEtran}

\usepackage{amsmath,amsfonts}
\usepackage{algorithmic}
\usepackage{array}
\usepackage{textcomp}
\usepackage{stfloats}
\usepackage{url}
\usepackage{verbatim}
\usepackage{graphicx}
\usepackage{balance}
\usepackage{cite}
\usepackage{tabularx}
\usepackage{amsmath} 
\usepackage{subcaption} 
\usepackage{booktabs}
\usepackage{cite} 
\usepackage{xcolor}

\title{Modeling Aesthetic Preferences in 3D Shapes: A Large-Scale Paired Comparison Study Across Object Categories}
\author{Kapil Dev,~\IEEEmembership{Member,~IEEE}%
\author{Kapil Dev\\
RMIT University, Melbourne, Australia\\
\texttt{kapil.dev@rmit.edu.vn}
}
\thanks{Kapil Dev is with RMIT University. Email: <your-email@domain>}
}

\begin{document}
\maketitle

\begin{abstract}
Human aesthetic preferences for 3D shapes are central to industrial design, 
virtual reality, and consumer product development. 
However, most computational models of 3D aesthetics lack empirical grounding 
in large-scale human judgments, 
limiting their practical relevance. 
We present a large-scale study of human preferences. 
We collected 22,301 pairwise comparisons across five object categories 
(chairs, tables, mugs, lamps, and dining chairs) via Amazon Mechanical Turk. 
Building on a previously published dataset~\cite{dev2020learning}, 
we introduce new non-linear modeling and cross-category analysis to uncover 
the geometric drivers of aesthetic preference.
We apply the Bradley-Terry model to infer latent aesthetic scores and use Random 
Forests with SHAP analysis to identify and interpret the most influential 
geometric features (e.g., symmetry, curvature, compactness). Our cross-category 
analysis reveals both universal principles and domain-specific trends in aesthetic preferences.
We focus on human-interpretable geometric features to ensure model 
transparency and actionable design insights, 
rather than relying on black-box deep learning approaches. 
Our findings bridge computational aesthetics and cognitive science, 
providing practical guidance for designers and a publicly available dataset 
to support reproducibility. 
This work advances the understanding of 3D shape aesthetics through a human-centric, data-driven framework.
\end{abstract}


\section{Introduction}
Despite the centrality of 3D shape aesthetics in design, virtual environments, 
and product development, there is a striking lack of computational models that 
are empirically grounded in large-scale human judgments. 
Most existing approaches rely on designer intuition or small-scale, 
subjective ratings, leaving open the fundamental question: 
\textit{Specifically, we ask which geometric properties truly drive human preferences for 3D shapes, 
and do these principles generalize across object categories?} Addressing 
this gap is critical for advancing both the science of aesthetics and practical 
applications in computer graphics and design automation.

\noindent
\textbf{This work addresses the following research questions:}
\begin{itemize}
    \item What geometric properties most strongly drive human preferences for 3D shapes?
    \item Are these aesthetic principles universal across object categories, or are they category-specific?
    \item Can non-linear modeling approaches reveal drivers of preference missed by linear analysis?
\end{itemize}

Several key gaps hinder the advancement of computational 3D aesthetics. Firstly, 
the reliance on absolute ratings like Likert scales introduces potential biases, 
whereas relational judgments obtained through paired comparisons offer a more robust 
method for eliciting preferences. Secondly, geometric properties unique to 3D shapes, 
such as symmetry, curvature, and compactness, which are crucial to aesthetic perception, 
lack systematic investigation. Finally, it remains unclear whether aesthetic principles 
are universal across different object categories, like chairs and lamps, 
or if they are specific to particular domains.

To address these gaps, this work introduces a novel dataset comprising 
22,301 pairwise comparisons of 3D shapes across five distinct categories, 
collected through Amazon Mechanical Turk. This data was then used to develop 
preference models based on the Bradley-Terry framework, allowing 
for the inference of aesthetic scores and their correlation with specific geometric 
features such as symmetry and curvature. Furthermore, the analysis of this data has 
yielded insights into cross-category aesthetics, identifying both universal principles, 
such as the importance of compactness, and category-specific drivers, 
like proportionality in chairs.

The dataset used in this study originates from our previous work~\cite{dev2020learning},  
this paper introduces a comprehensive non-linear modeling framework (Random Forests), 
interprets feature importances using SHAP, and rigorously tests the consistency of aesthetic 
principles across object categories—analyses not explored in the original dataset paper.

In summary, our work advances computational aesthetics by providing a robust, 
interpretable framework for modeling and understanding human preferences for 3D shapes.
This work offers valuable new perspectives on the complex relationship 
between the form of an object and its aesthetic perception, as well as 
the interplay between form and function.
Unlike recent deep learning approaches that learn abstract features, 
we deliberately restrict our analysis to human-interpretable geometric descriptors, 
enabling transparent insights on human-understandable shape properties.
\textit{Crucially, our analysis reveals that simple linear 
relationships between geometric features and perceived aesthetics are 
often insufficient, necessitating the use of models capable of capturing 
non-linear interactions to accurately identify the true drivers of preference.}

\vspace{2em} 

\noindent \textbf{Our Contributions:}
\begin{itemize}
    \item We introduce a large-scale paired comparison dataset for 3D shape aesthetics 
    across five distinct object categories.
    \item We apply the Bradley-Terry model to infer latent aesthetic scores from human judgments, 
    providing a robust quantitative framework for preference analysis.
    \item We develop a comprehensive feature extraction pipeline that integrates geometric, 
    structural, and curvature-based descriptors to capture the key drivers of aesthetic appeal.
    \item We leverage non-linear modeling techniques, including Random Forests and SHAP, to uncover 
    both universal and category-specific aesthetic principles.
    \item We offer actionable insights for design optimization and lay the groundwork for future 
    interdisciplinary research in computational aesthetics.
\end{itemize}


\section{Related Work}
The computational modeling of 3D shape aesthetics intersects computer graphics, 
cognitive science, and design research. Prior work can be broadly categorized into 
three areas: (1) computational aesthetics in 3D, (2) empirical studies of human 
preferences, and (3) datasets and methodologies for large-scale analysis.

Despite significant progress, several gaps remain. 
Many computational approaches rely on small-scale or synthetic datasets, 
limiting their generalizability. Empirical studies often lack diversity in 
object categories or use subjective rating scales prone to bias. 
Furthermore, while deep learning methods have advanced feature discovery, 
their lack of interpretability hinders actionable design insights. 
Our work addresses these gaps by leveraging large-scale, human-annotated 
paired comparisons and focusing on interpretable geometric features across 
multiple categories.

\subsection{Computational Aesthetics in 3D}
The field of computational aesthetics emerged with the goal of developing computational methods 
capable of making aesthetic decisions in a manner that aligns with human perception \cite{neumann2005defining}.
Early efforts in this domain sought to formalize and quantify aesthetic principles.

Over time, the field of computational aesthetics, particularly concerning 3D shapes, has 
undergone a significant transformation, shifting towards data-driven methodologies that harness 
the power of machine learning and statistical analysis of human judgments \cite{dev2020learning}.
These methods often employ neural networks to process raw 3D shape representations 
(e.g., voxel grids, point clouds, or multi-view images). This enables the discovery of aesthetic 
features without explicit manual definition, but often at the cost of interpretability.
Beyond the general definitions of computational aesthetics, specific research has focused on 
defining and computationally measuring the artistic perception and interestingness of 3D shapes 
\cite{lau2023computational}. Empirical studies highlight symmetry and geometric properties as 
key factors influencing human aesthetic preferences for 3D shapes \cite{jayadevan2018perception, pizlo2021concept}.
However, deep learning approaches, while powerful, often lack transparency and do not 
provide direct insight into which geometric properties drive human preferences. 
This motivates the need for interpretable models that can inform both theory and practice in design.

\subsection{Empirical Studies and Human-Centric Datasets}
Empirical investigations of 3D aesthetics remain limited in scale and diversity. 
ShapeNet \cite{chang2015shapenet}, a canonical dataset of 3D CAD models, has fueled research on shape 
classification and reconstruction but lacks standardized aesthetic annotations. 
Given the subjective nature of aesthetic appreciation, 
computational models aiming to predict or understand aesthetic preferences for 3D shapes 
must be firmly grounded in empirical data derived from human evaluations \cite{dev2023comparing}.
Various methodologies are employed to elicit these judgments, including 
the use of rating scales where participants assign a score to a shape based on its aesthetic appeal, 
ranking tasks where participants order a set of shapes according to their preferences, 
and, particularly relevant to the current study, pairwise comparisons where participants 
indicate which of two presented shapes they find more aesthetically pleasing \cite{dev2020learning}.
Interestingly, research has explored the influence of shape representation on aesthetic judgments, 
with findings suggesting that humans can often make aesthetic decisions even when presented with 
relatively coarse representations of the shapes, such as low-resolution point clouds or 
voxelizations \cite{dev2023comparing}.

The progress in data-driven research on computational 3D aesthetics has been significantly 
supported by the creation of several human-centric datasets. One notable example is the ViDA 3D 
dataset, which comprises a large collection of 3D models 
sourced from the online platform Sketchfab \cite{angelini2020vida}.
Additionally, several datasets have been specifically curated for pairwise comparison studies, 
where human participants directly compare the aesthetics of two or more shapes \cite{dev2020learning}.
Friedenberg \cite{friedenberg2012aesthetic} demonstrated that perceived attractiveness of 
triangles is driven by compactness (axis ratio) rather than the golden ratio, 
with upward-pointing orientations deemed more stable and appealing.

\subsection{Advances in Preference Modeling}
Pairwise comparison method is often considered less cognitively demanding for participants 
compared to rating scales and directly captures relative preferences \cite{dev2020learning}.
A powerful statistical framework for analyzing the outcomes of such studies is the Bradley-Terry 
model \cite{bradley1952rank}. This model allows for the inference of latent aesthetic scores 
for each shape in the dataset based on the pattern of wins and losses in the pairwise comparisons.
The application of deep learning-to-rank algorithms aims to develop a scoring function that can rank 3D 
shapes according to their aesthetic appeal based on the provided preference judgments \cite{dev2020learning}.

\subsection{3D Shape Analysis and Feature Extraction for Aesthetic}
A fundamental step in computationally modeling aesthetic preferences for 3D shapes involves 
analyzing and extracting relevant features that can capture the essence of a shape's form 
\cite{chen2022learning, kazmi2013survey}. 
Among these, geometric descriptors play a crucial role. These features encompass a wide range of 
properties, including symmetry (such as bilateral or rotational symmetry), curvature (quantifying 
how much a surface bends, e.g., mean and Gaussian curvature), aspect ratios (describing the proportional 
relationships of a shape's dimensions), volume, and surface area 
\cite{matsuyama2023systematic, kazmi2013survey, das2017local}. These basic geometric properties 
often serve as intuitive and interpretable correlates of aesthetic judgments, as they capture fundamental 
aspects of a shape's form that might influence human aesthetic perception. Their relative ease of 
computation makes them a common starting point for aesthetic analysis \cite{kazmi2013survey, lara2017comparative}.
Beyond the overall geometric form, the structural organization of a 3D shape, 
including the identification of its constituent parts and their spatial relationships, 
is also critical for understanding aesthetic preferences \cite{mitra2014structure}

In our work, we build on these established methodologies by integrating a comprehensive feature computation pipeline with 
non-linear analysis tools to derive robust aesthetic preference scores from paired comparison data. This approach not only 
validates earlier studies but also refines the understanding of perceptual cues in 3D shape aesthetics.

In summary, while prior work has advanced both computational and empirical approaches to 3D shape aesthetics, 
key challenges remain: limited dataset diversity, reliance on subjective or absolute ratings, 
and a lack of interpretable models. Our study addresses these by leveraging large-scale, 
category-diverse paired comparisons and focusing on transparent, geometric feature-based 
modeling to uncover both universal and category-specific drivers of aesthetic preference.

\section{Methods}
This section outlines our methodological pipeline, covering dataset construction, 
pairwise preference collection, strategies for addressing dataset imbalance, 
modeling of latent aesthetic scores, and statistical analyses to interpret 
feature importance and cross-category consistency.

\subsection{Dataset Construction}
We adopt the dataset from our prior work \cite{dev2020learning}, 
which collected human aesthetic judgments for 3D shapes across five categories: 
club chairs (778 models), dining chairs (277), lamps (78), mugs (65), and tables (30). 
Below, we summarize key aspects of the dataset; 
full methodological details (e.g., crowdsourcing protocols, preprocessing) are described in \cite{dev2020learning}. 
3D models were centered, category-scaled to preserve proportions, 
and had vertex normals recomputed for consistent surface features (Table~\ref{tab:dataset_summary}).

\begin{table}[h!]
      \centering
      \caption{Summary of Pairwise Comparisons}
      \begin{tabular}{lcc}
        \hline
        Category & \# Models & \# Comparisons \\
        \hline
        Club chairs & 778 & 9,875 \\
        Dining chairs & 277 & 5,726 \\
        Lamps & 78 & 3,250 \\
        Mugs & 65 & 1,000 \\
        Tables & 30 & 3,250 \\ \hline
        Total & 1,248 & 22,301 \\
        \hline
      \end{tabular}
      \label{tab:dataset_summary}
\end{table}

\noindent
Pairwise comparisons were generated by randomly 
sampling pairs of shapes within each category. 
Attention-check pairs and exclusion criteria followed the protocol in \cite{dev2020learning}.

We acknowledge the variation in model 
counts per category, 
a limitation inherent in the original dataset \cite{dev2020learning}. 
As Table~\ref{tab:dataset_summary} shows, model counts vary widely across categories 
(e.g., 778 club chairs vs.\ 30 tables). To ensure this imbalance did not bias our inferences, 
we adopted two strategies:
\begin{itemize}
  \item \textbf{Per-category fitting:} All Bradley-Terry and Random Forest models were trained 
  separately per category, so no category ``dominates'' another in a pooled fit.
  \item \textbf{Sample weighting:} Within each Random Forest, we assigned each comparison a weight 
  inversely proportional to the number of comparisons for its shape. Specifically, 
  for a shape $i$ with $n_i$ comparisons, the weight $w_i = 1/n_i$ was used. These weights were passed 
  to the \texttt{sample\_weight} parameter in scikit-learn's \texttt{RandomForestRegressor}.
  \end{itemize}

\subsection{Bradley-Terry Model}
The Bradley-Terry model is a probabilistic framework 
widely used to analyze pairwise comparison data. 
In this work, we leverage the Bradley-Terry model to 
infer latent aesthetic scores for 3D shapes based on human judgments. 

The model assumes that each item (e.g., a 3D shape) is associated with a 
latent score, which reflects its relative preference in pairwise comparisons. 

Given two items \(i\) and \(j\), the probability that 
item \(i\) is preferred over item \(j\) is modeled as:

\[
P(i \succ j) = \frac{\exp(\beta_i)}{\exp(\beta_i) + \exp(\beta_j)},
\]

where \(\beta_i\) and \(\beta_j\) are the latent scores for items \(i\) and \(j\), respectively.

To estimate these scores, we maximize the likelihood of the observed pairwise comparison data. 
Specifically, for a dataset of \(N\) comparisons, 
where each comparison indicates whether item \(i\) was preferred over item \(j\), 
the likelihood function is given by:

\[
\mathcal{L}(\boldsymbol{\beta}) = \prod_{k=1}^{N} P(i_k \succ j_k)^{y_k} \cdot P(j_k \succ i_k)^{1-y_k},
\]

where \(y_k = 1\) if item \(i_k\) is preferred over \(j_k\), and \(y_k = 0\) otherwise. 
The parameters \(\boldsymbol{\beta}\) are estimated using maximum likelihood estimation (MLE), 
with regularization applied to ensure numerical stability and prevent overfitting.

In our study, we applied the Bradley-Terry model to pairwise comparison data 
collected via Amazon Mechanical Turk. The dataset spans five object categories—chairs, 
tables, mugs, lamps, and dining chairs—comprising over 22,000 comparisons. 
For each category, we constructed a design matrix encoding the pairwise relationships 
and used logistic regression to fit the model. 
The resulting latent scores provide a quantitative measure of aesthetic preference for each shape.

The inferred scores were further analyzed to identify geometric features driving aesthetic preferences. 
By correlating the latent scores with features such as symmetry, curvature, and compactness, 
we uncovered both universal principles and category-specific trends in 3D shape aesthetics. 
These insights form the basis for subsequent statistical analyses and feature 
importance evaluations presented in this work.

The Bradley-Terry model was implemented using the \texttt{statsmodels} Python package, with L2 
regularization ($\lambda=0.01$) to ensure numerical stability. For each category, a design 
matrix was constructed from the pairwise comparison files. 
Model convergence was monitored via log-likelihood and gradient norms. 
Standard errors for the latent scores were estimated using the Fisher information matrix. 
All code and scripts for fitting and diagnostics will be made available on GitHub.

\begin{figure*}[htb!]
  \centering
  \includegraphics[width=.27\linewidth]{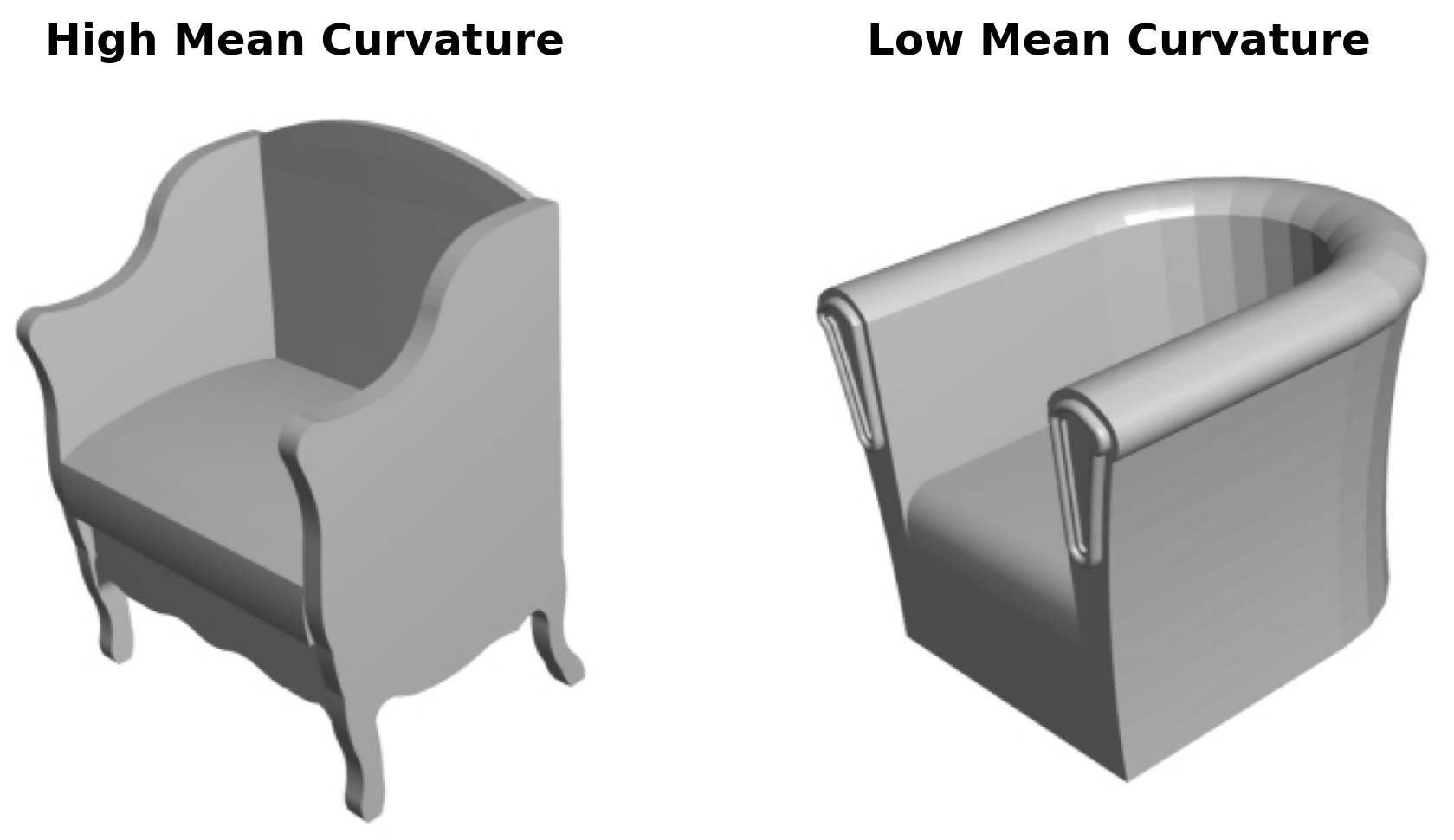}
  \hspace{1.5em}
  \vrule width 1.0pt
  \hspace{1.5em}
  \includegraphics[width=.27\linewidth]{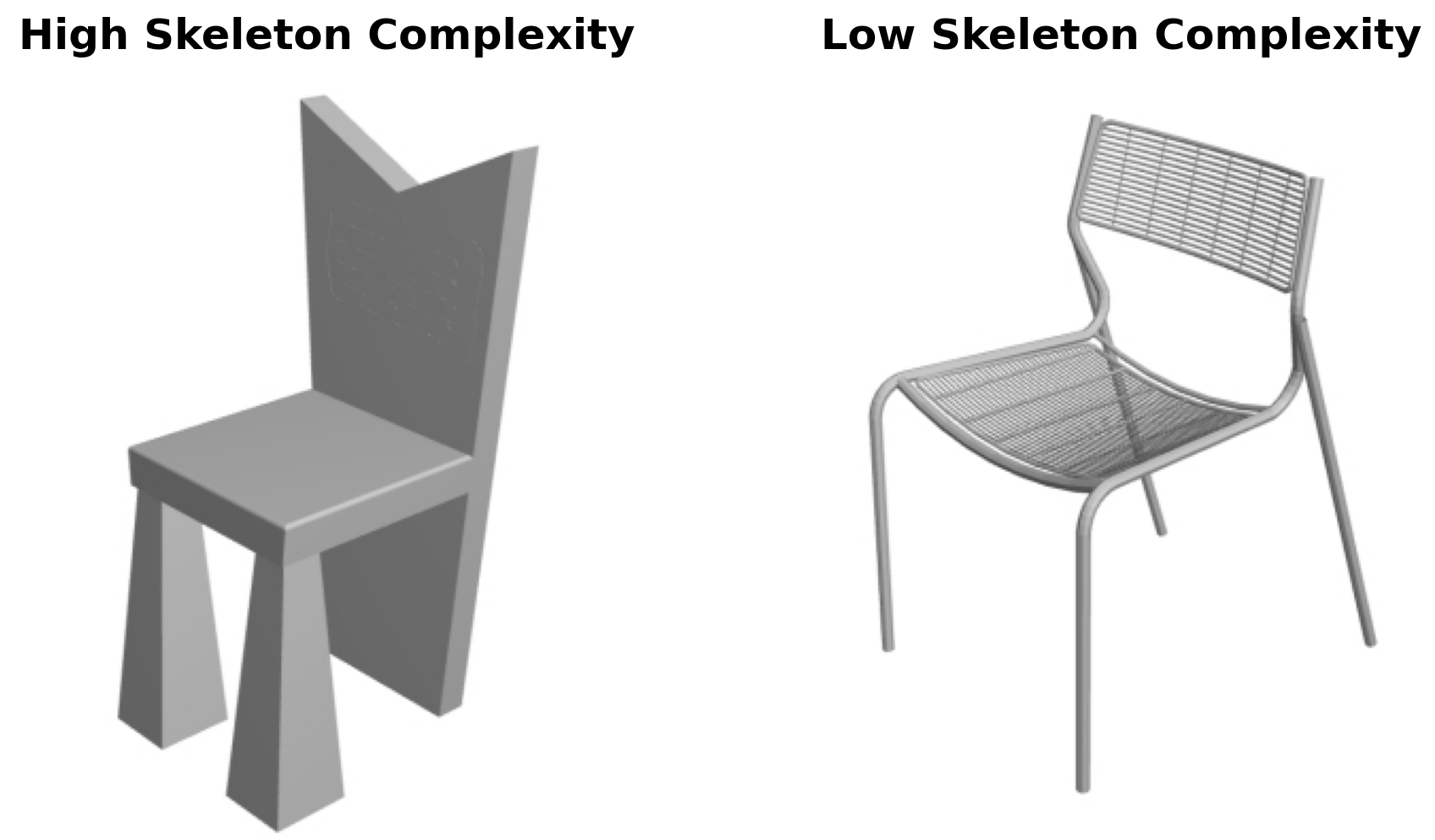}
  \hspace{1.5em}
  \vrule width 1.0pt
  \hspace{1.5em}
  \includegraphics[width=.27\linewidth]{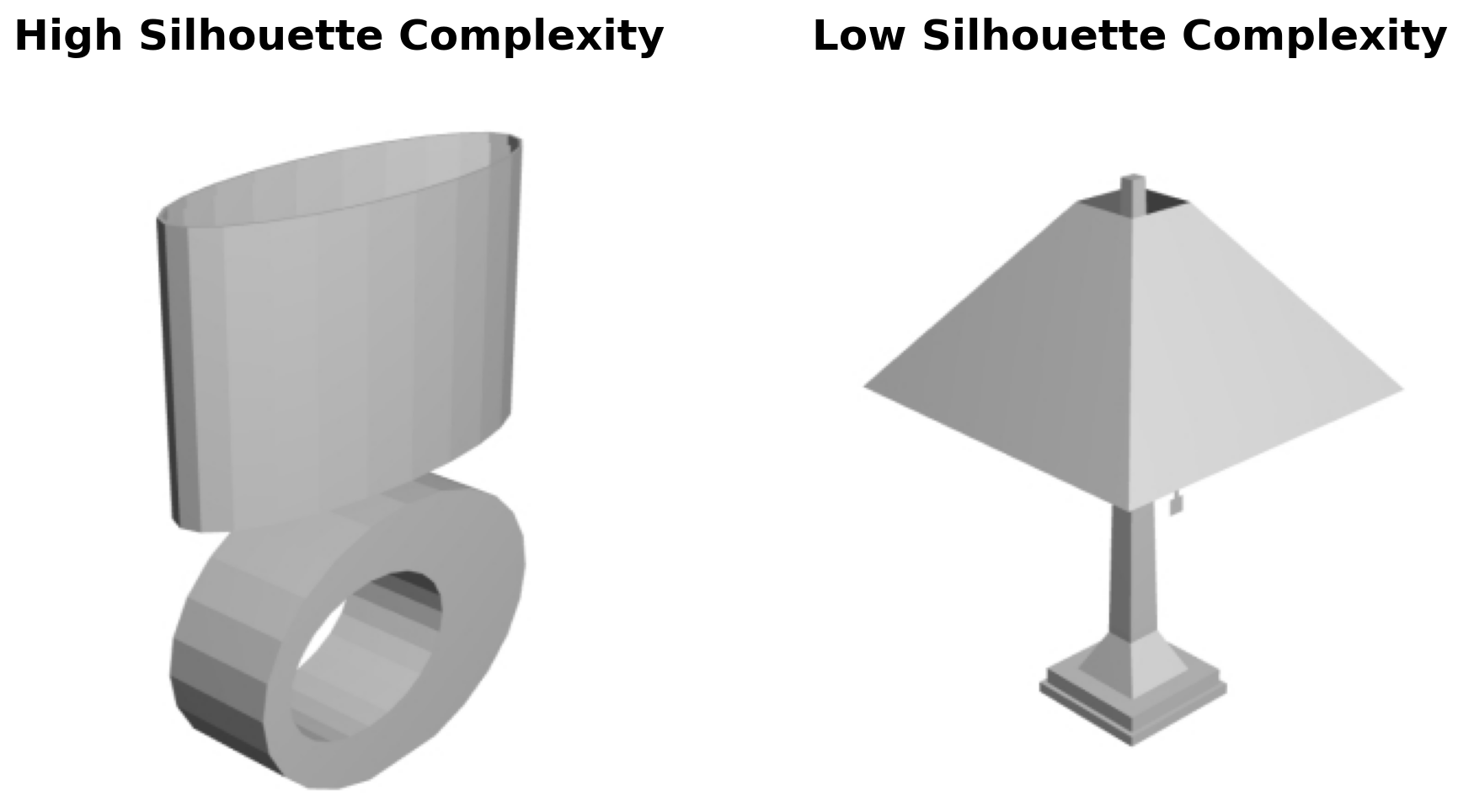}
  \caption{\label{fig:keyFeatures}
  Representative 3D shapes illustrating high and low values of key geometric features: 
  mean curvature (club chairs), 
  skeleton complexity (dining chairs), 
  and silhouette complexity (lamps). 
  These examples visualize the types of shapes associated with different ends of 
  the aesthetic score distributions in each category.}
\end{figure*}

\subsection{Feature Extraction}
The feature computation pipeline extracts a diverse set of geometric, structural, 
and curvature-based features from both the mesh and point cloud representations 
of 3D shapes. All features were selected for their interpretability, 
allowing direct mapping between model predictions and human-understandable shape properties.

These features are designed to capture the essential 
characteristics of the shapes that influence perceptual aesthetics (see Figure~\ref{fig:keyFeatures} and Figure~\ref{fig:feature_distributions}):

\begin{figure}[ht]
    \centering
    \includegraphics[width=0.49\linewidth]{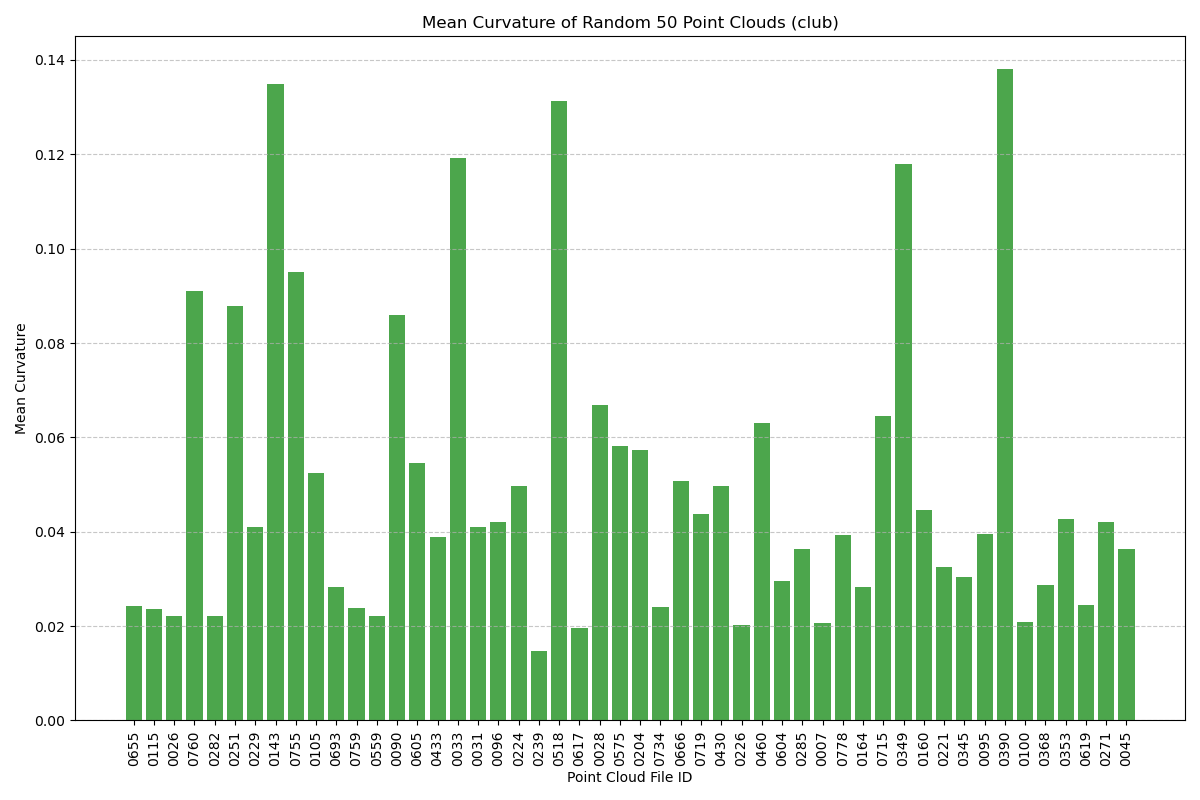}
    \hfill
    \includegraphics[width=0.49\linewidth]{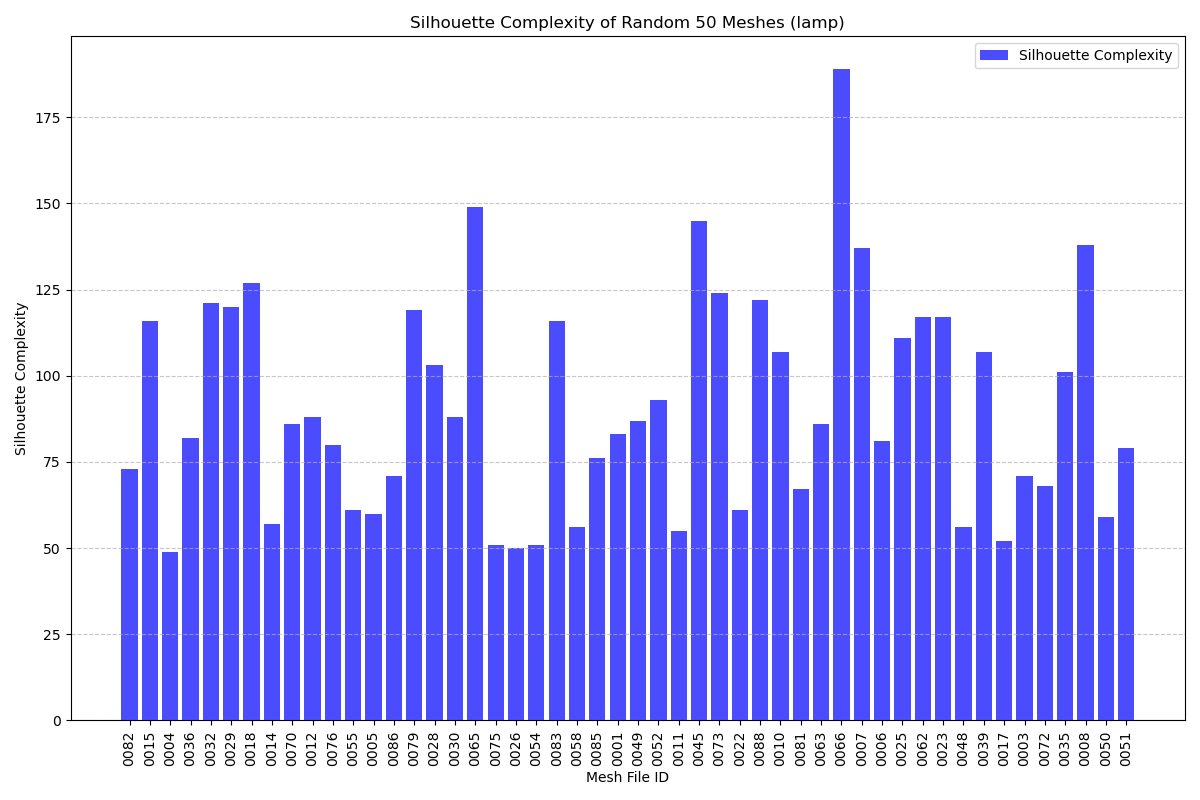}
    \caption{Example feature distributions for individual categories. Left: Mean Curvature for the club category. Right: Silhouette Complexity for the lamp category. These plots highlight the variation of key features across shapes within each category.}
    \label{fig:feature_distributions}
\end{figure}

\textbf{Curvature Features}
Curvature features are computed from the point cloud representation of the shape using Principal Component Analysis (PCA) on local neighbourhood. These features include:
\begin{enumerate}
    \item Mean Curvature: The average curvature across the shape's surface.
    \item Curvature Variance: The variability in curvature, capturing surface irregularities.
    \item Median Curvature: A robust measure of the central tendency of curvature values.
\end{enumerate}
These features quantify the smoothness and variability of the shape's surface, which are critical for understanding its ergonomic and aesthetic properties.

\textbf{Shape Compactness}
Compactness features are derived from the mesh representation and include:
\begin{enumerate}
    \item Surface-to-Volume Ratio: The ratio of the surface area to the volume of the shape, which quantifies its compactness.
    \item Convexity Ratio: The ratio of the shape's volume to the volume of its convex hull, providing a measure of how closely the shape approximates a convex form.
\end{enumerate}
These features are particularly relevant for capturing the overall compactness and structural efficiency of the shape.

\textbf{Proportional Features}
Proportional features are computed from the oriented bounding box of the mesh and include:
\begin{enumerate}
    \item Aspect Ratio X: The ratio of the extents along the X-axis to the Y-axis.
    \item Aspect Ratio Y: The ratio of the extents along the Y-axis to the Z-axis.
    \item Aspect Ratio Z: The ratio of the extents along the Z-axis to the X-axis.
\end{enumerate}
These features capture the relative dimensions of the shape, which are critical 
for defining its proportions and visual balance.

\textbf{Silhouette and Contour Features}
Silhouette and contour features are derived from the mesh and include:
\begin{enumerate}
    \item Silhouette Complexity: The complexity of the shape's silhouette, computed by 
    slicing the mesh along multiple planes and counting the number of silhouette edges.
    \item Multi-View Silhouette Complexity: The total silhouette complexity across multiple views, 
    obtained by rotating the mesh around the Z-axis and analyzing its silhouette from different angles.
    \item Hollow Ratio: The ratio of the volume difference between the convex hull and the actual 
    shape to the convex hull volume, capturing the degree of hollowness.
\end{enumerate}
These features provide insights into the shape's outline and structural intricacy, 
which are important for aesthetic evaluation.

\textbf{Skeleton Complexity}
Skeleton complexity is computed by converting the mesh into a voxel grid, applying morphological 
thinning, and measuring the size of the resulting skeleton. This feature quantifies the structural 
intricacy of the shape, which is particularly relevant for categories like lamps and tables.

\begin{table}[h!]
\footnotesize
\setlength{\tabcolsep}{3pt}
\renewcommand{\arraystretch}{1.05}
\caption{Summary of Extracted Geometric Features}
\begin{tabularx}{\columnwidth}{lX}
\toprule
\textbf{Feature} & \textbf{Description} \\
\midrule
Mean Curvature & Average curvature across the surface \\
Curvature Variance & Variability in surface curvature \\
Median Curvature & Median of curvature values \\
Surface-to-Volume Ratio & Surface area divided by volume (compactness) \\
Convexity Ratio & Volume divided by convex hull volume \\
Aspect Ratio X, Y, Z & Ratios of bounding box extents along axes \\
Silhouette Complexity & Edge count of silhouette (single view) \\
Multi-View Silhouette Complexity & Total silhouette complexity across multiple views \\
Hollow Ratio & (Convex hull vol. $-$ shape vol.) / convex hull vol. \\
Skeleton Complexity & Size of skeleton after voxelization and thinning \\
\bottomrule
\end{tabularx}
\label{tab:feature_summary}
\end{table}

All geometric features were computed using the \texttt{trimesh} and \texttt{Open3D} 
Python libraries. Skeleton complexity was computed by voxelizing each mesh at a 
resolution of $64^3$ and applying the Zhang-Suen thinning algorithm. 
Curvature features were estimated using PCA on local neighborhoods of 20 points. 
 
\subsection{Statistical Analysis}
\subsubsection{Linking Geometric Features to Aesthetic Scores}
As an initial exploratory step, we computed Pearson correlation 
coefficients between each extracted geometric feature and the derived 
BT latent scores to assess basic linear trends (see Section~\ref{sec:ResultsCorrelation}). However, to capture potentially complex, 
non-linear relationships and feature interactions influencing aesthetic preferences, 
we trained a Random Forest regression model to predict Bradley-Terry (BT) 
latent scores (interpreted as aesthetic preference scores) from extracted 
geometric features.

\noindent \textbf{Model Setup:} The model was designed to predict BT 
aesthetic scores (a continuous variable) based on geometric features 
such as symmetry, curvature, and compactness. Hyperparameters, 
including the number of trees and maximum depth, were optimized using a 
grid search with 5-fold cross-validation to prevent overfitting. 
The model's performance was evaluated using $R^2$ (R-squared), 
Mean Absolute Error (MAE), and out-of-bag (OOB) error.

\noindent \textbf{Interpretability:} To understand the model's 
decision-making process, we employed SHAP (SHapley Additive exPlanations) 
to compute Shapley values, quantifying the contribution of each feature to 
the predictions. This approach allowed us to identify globally important 
features, such as curvature, and reveal interactions between features, 
like symmetry $\times$ proportionality. 

Additionally, we utilized Partial Dependence Plots (PDPs) to visualize 
the marginal effects of individual features on the BT aesthetic scores, 
illustrating how aesthetic preference changes as features like compactness 
increase. For each category, PDPs were generated for the 
top-ranked features identified by SHAP analysis. This approach enables 
visualization of non-linear and interaction effects that are not apparent 
from linear correlation coefficients.

\subsubsection{Cross-Category Consistency Tests}
To assess whether aesthetic principles generalize across categories 
(e.g., chairs vs. lamps), we conducted three analyses:

\noindent \textbf{Feature Importance Correlation:} To assess the consistency 
of feature importance across different categories, we computed Spearman's rank 
correlation between SHAP-derived feature importances for pairs of categories 
(e.g., chairs vs. tables). A high correlation indicates shared drivers of aesthetic 
preference across these categories (e.g., compactness). 
The pairwise correlations were visualized using a heatmap (Figure 5).

\noindent \textbf{Model Transferability:} To evaluate the generalizability of the 
learned aesthetic principles, we assessed model transferability. 
This involved training a Random Forest model on one category (e.g., chairs) and testing 
it on another (e.g., lamps) to quantify the performance drop. A small drop in $R^2$ 
suggests the presence of universal aesthetic principles, while a large drop implies 
category-specific aesthetic preferences.

\noindent \textbf{Cluster Analysis of Feature Spaces:} Hierarchical clustering was applied 
to the feature importance vectors (calculated per category) to group categories that 
exhibit similar aesthetic drivers. This analysis aimed to identify clusters of categories 
sharing underlying principles of aesthetic preference.

Random Forest regressors were implemented using scikit-learn 
(\texttt{RandomForestRegressor}), with hyperparameters (number of trees, 
maximum depth, minimum samples per leaf) optimized via grid search and 5-fold 
cross-validation (random seed 42). SHAP values were computed using the 
TreeExplainer method, and global feature importance was aggregated as the 
mean absolute SHAP value across all samples. Partial dependence plots were 
generated using scikit-learn's \texttt{partial\_dependence} function.

All code for feature extraction, model fitting, and statistical analysis 
will be made publicly available on GitHub. The full dataset, including pairwise 
comparison files and latent BT scores, will be released upon publication.

\begin{figure}[htb!]
  \centering
  \includegraphics[width=.95\linewidth]{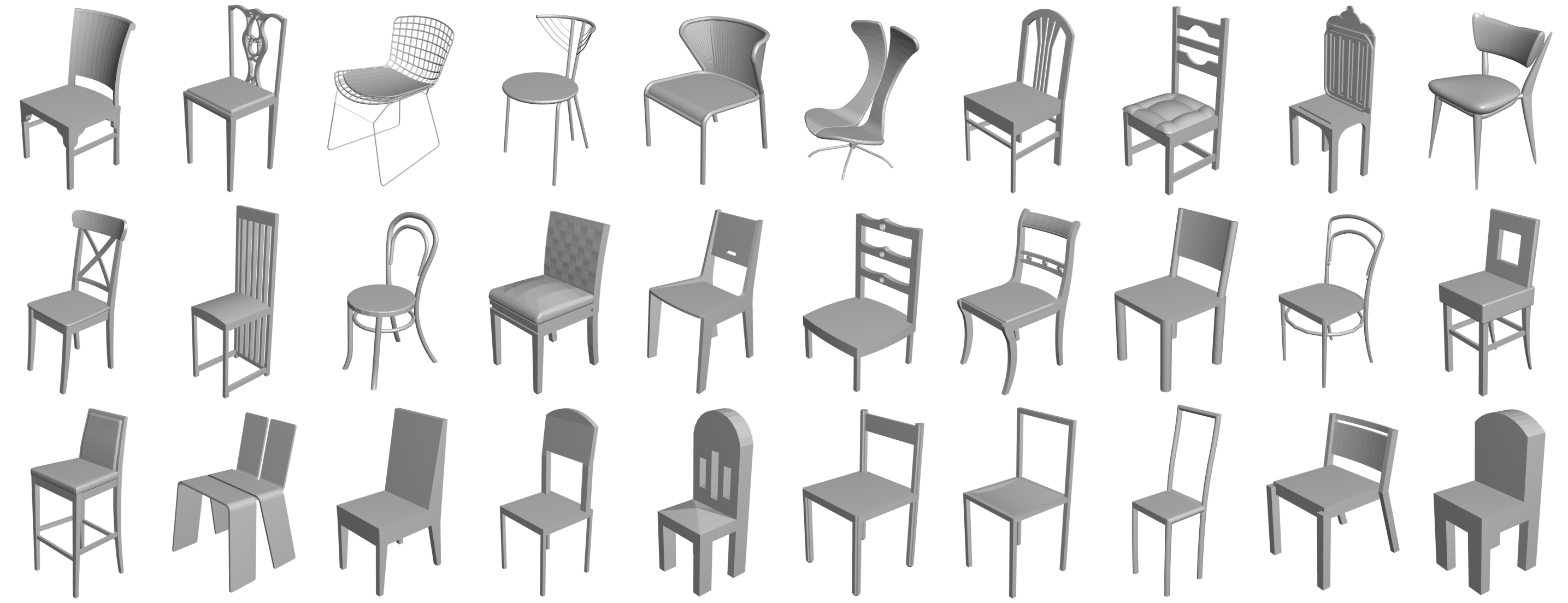}
  \textcolor{gray}{\rule{1\linewidth}{1pt}}
  \includegraphics[width=.95\linewidth]{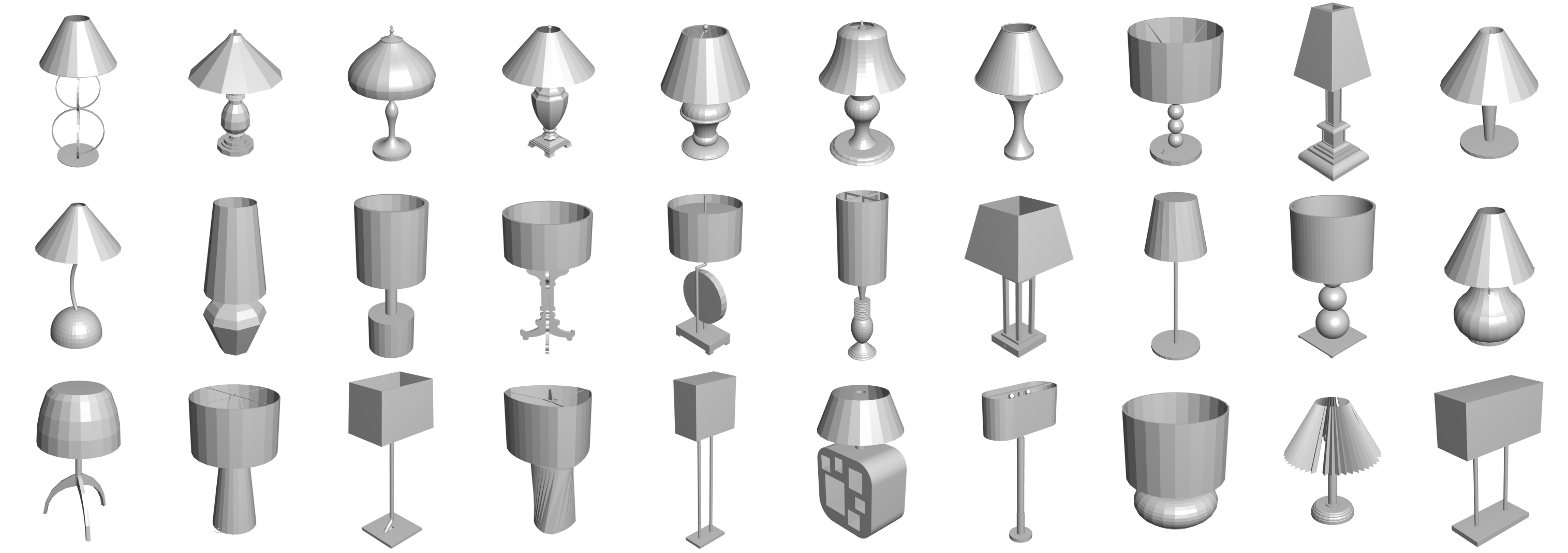}
  \textcolor{gray}{\rule{1\linewidth}{1pt}}
  \includegraphics[width=.98\linewidth]{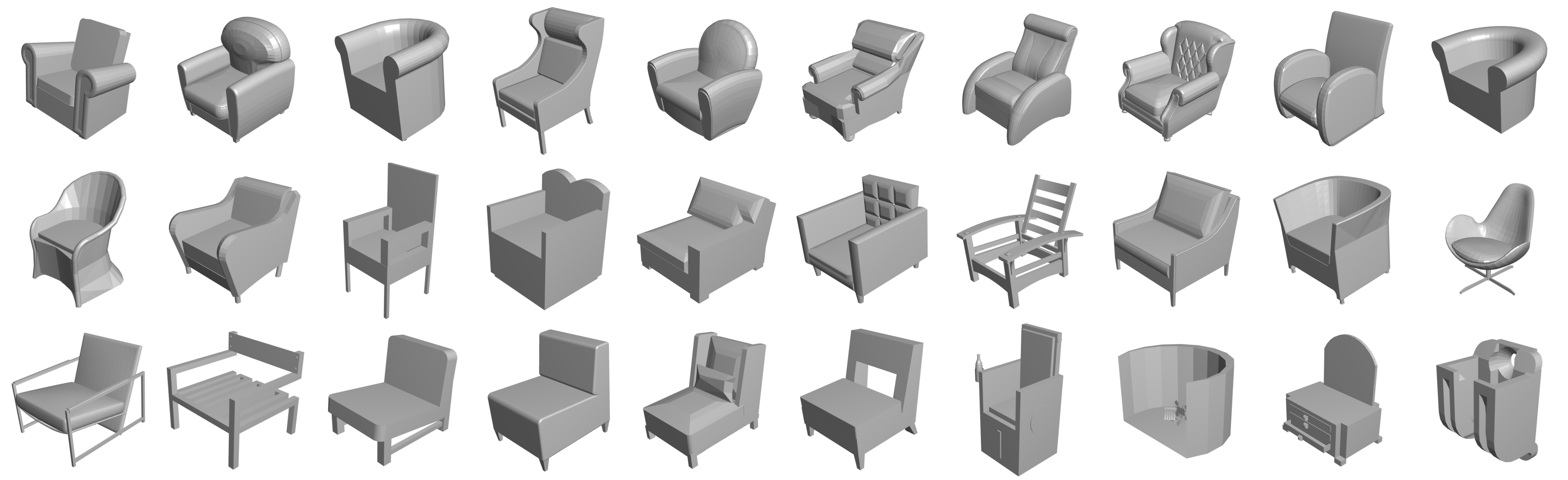}
  \textcolor{gray}{\rule{1\linewidth}{1pt}}
  \includegraphics[width=.92\linewidth]{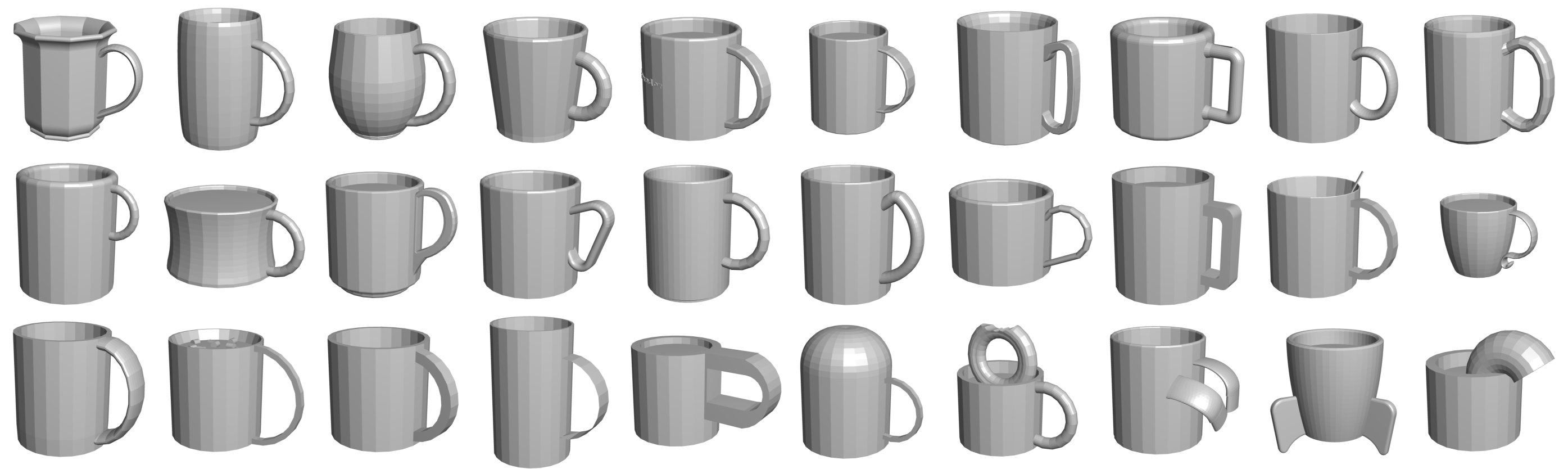}
  \caption{\label{fig:shapeRankings}
    Visualized Bradley-Terry aesthetic rankings for four categories: dining chairs, lamps, clubs and mugs. 
    For each category, the top row shows the 10 shapes with the highest inferred aesthetic scores, 
    the middle row shows 10 randomly selected shapes, and the bottom row shows the 10 shapes with 
    the lowest scores. Shapes are ordered from highest (left) to lowest (right) within each row, 
    illustrating the diversity of human preferences and geometric variation.
  }
\end{figure}


\section{Results}
\subsection{Inference of Aesthetic Rankings}
The Bradley-Terry (BT) model was employed to infer latent aesthetic scores 
for 3D shapes based on pairwise comparison data. 
These scores provide a quantitative measure of aesthetic preferences, 
enabling the ranking of shapes within each category. 
Figure~\ref{fig:shapeRankings} illustrates the visualized rankings 
for selected categories, highlighting the diversity in aesthetic 
appeal across shapes.

The BT model results reveal distinct trends in aesthetic preferences 
across categories. For instance, in the club chair and dining chair 
categories, shapes with balanced proportions and smooth curvature 
tend to rank higher, reflecting a preference for ergonomic and visually 
harmonious designs. Conversely, in the lamp category, structural intricacy, 
as captured by features like skeleton complexity and silhouette complexity, 
plays a more significant role in determining aesthetic appeal.

\subsection{Distribution of Inferred Aesthetic Scores by Category}

\begin{figure}[htb!]
      \centering
      \includegraphics[width=.49\linewidth]{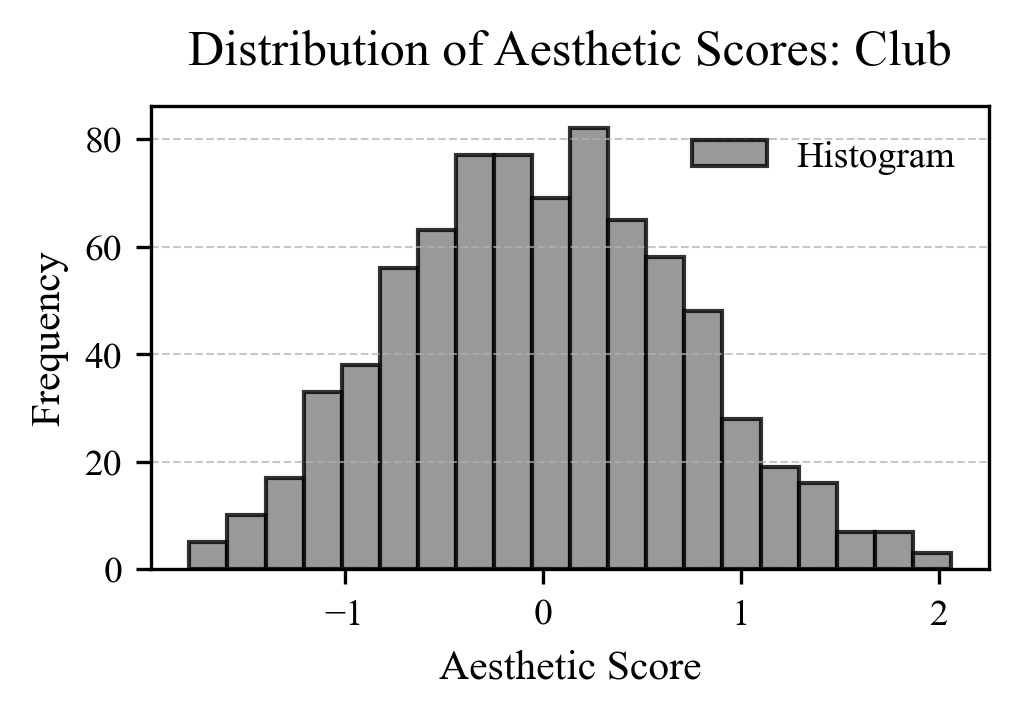}
      \hfill
      \includegraphics[width=.49\linewidth]{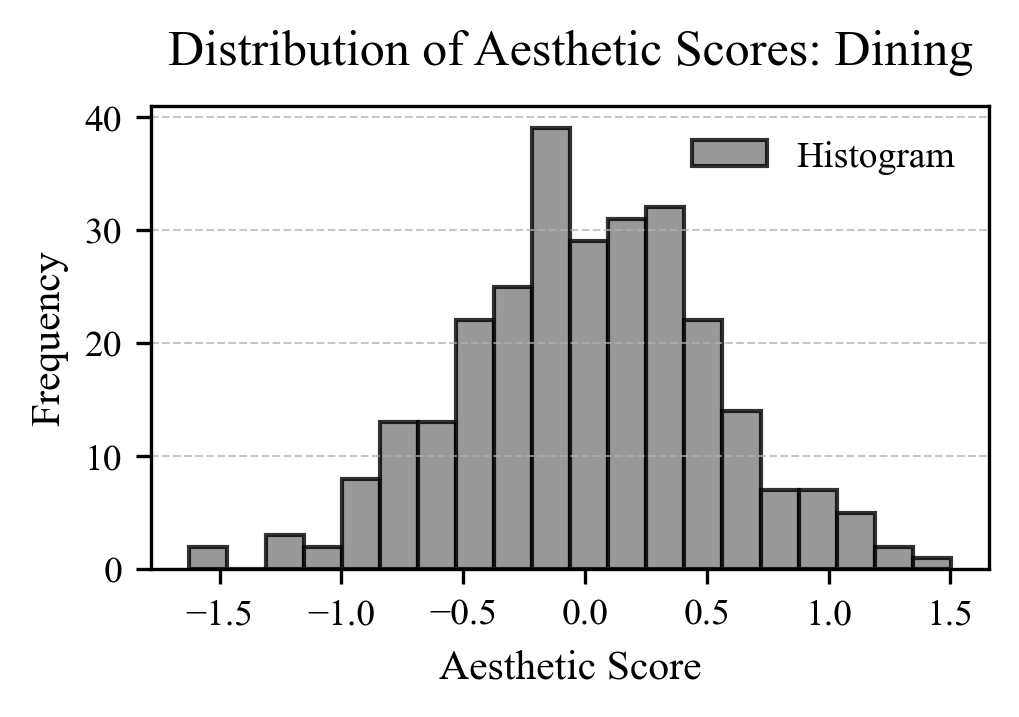}
      \hfill
      \includegraphics[width=.49\linewidth]{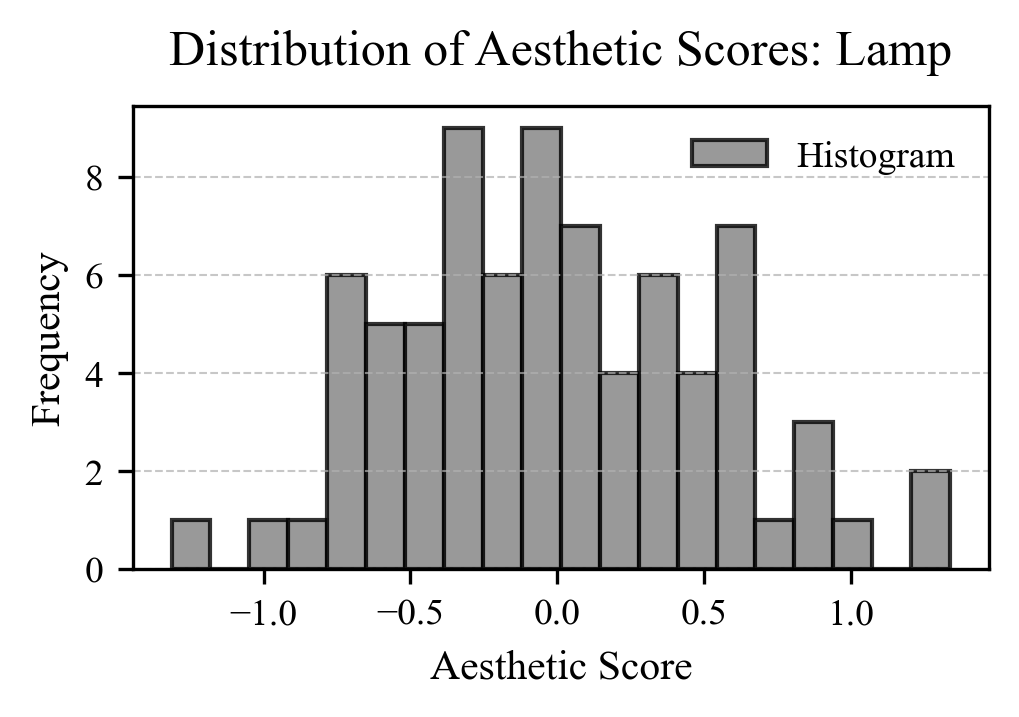}
      \hfill
      \includegraphics[width=.49\linewidth]{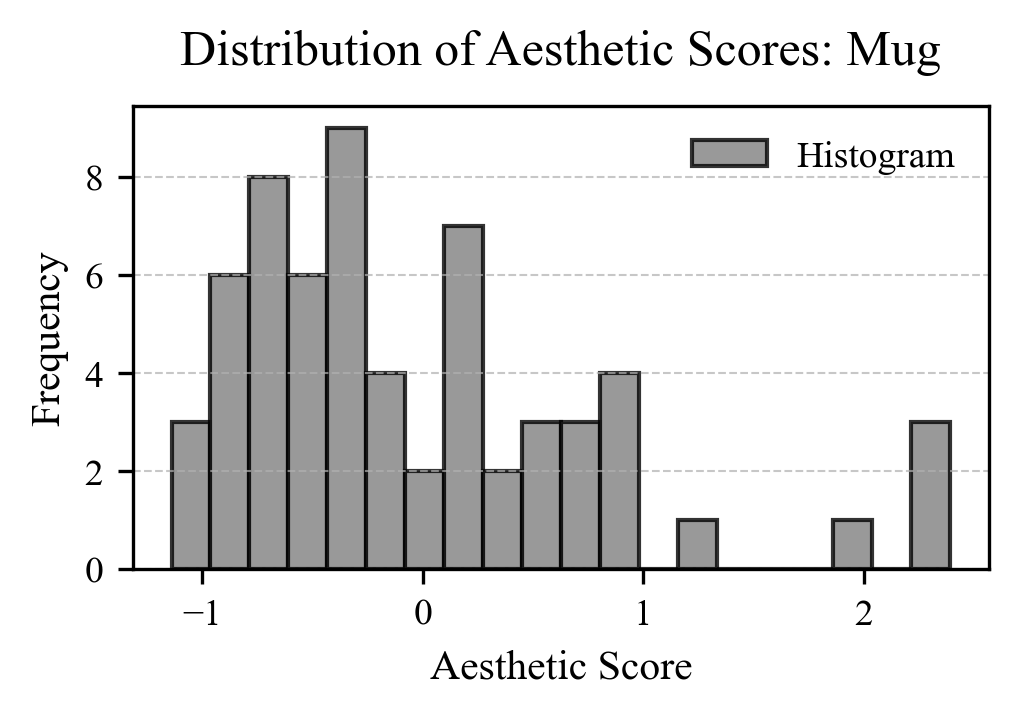}
      \caption{\label{fig:distribution}
                               Distribution of aesthetic scores: club chairs, 
                               dining chairs, lamps, and mugs.}
\end{figure}

Figure~\ref{fig:distribution} shows the distribution of inferred Bradley-Terry (BT) 
aesthetic scores for each category. Club and dining chairs exhibit symmetric, 
narrow distributions of inferred BT scores, 
suggesting greater agreement among participants within these categories. 
In contrast, lamps and mugs display broader or skewed distributions, indicating higher 
variability in aesthetic preferences for these categories.

\begin{table}[h!] 
      \centering
      \caption{Random Forest Model Performance Metrics for Predicting BT Scores. 
      R² and MAE are reported on the held-out test set, while OOB Error is the Out-of-Bag error estimate from the training data.}
      \begin{tabular}{lccc}
      \toprule 
      Category & R² Score & MAE & OOB Score \\
      \midrule 
      Club & 0.31 & 0.45 & 0.16 \\
      Dining & 0.13 & 0.41 & 0.05 \\
      Lamp & 0.11 & 0.45 & 0.08 \\
      Mug & 0.05 & 0.58 & 0.07 \\
      Table & 0.48 & 0.42 & 0.07 \\
      \bottomrule 
      \end{tabular}
\label{tab:model_performance}
\end{table}

\begin{figure}[htb!]
  \centering
  \includegraphics[width=.98\linewidth]{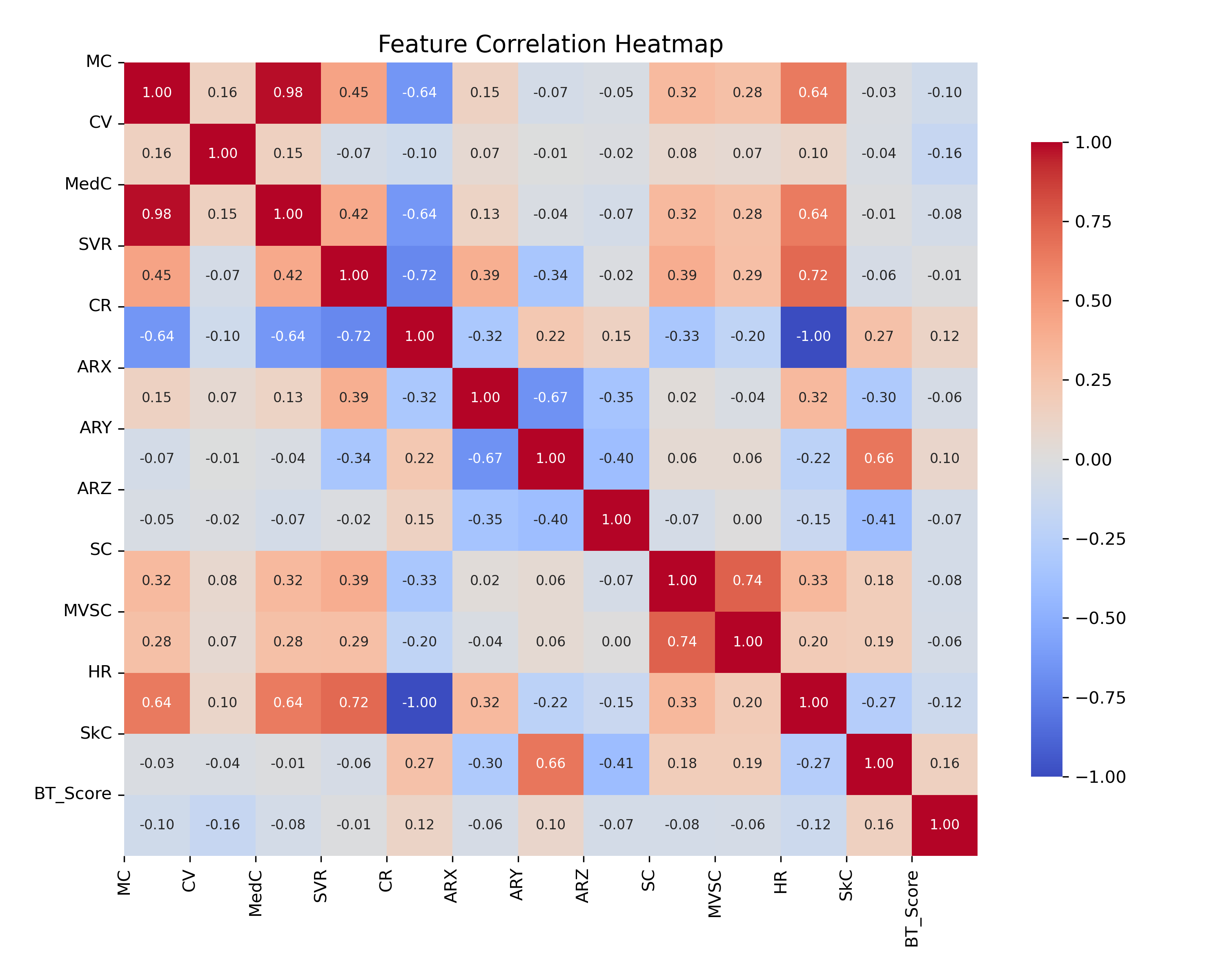}
  \includegraphics[width=.98\linewidth]{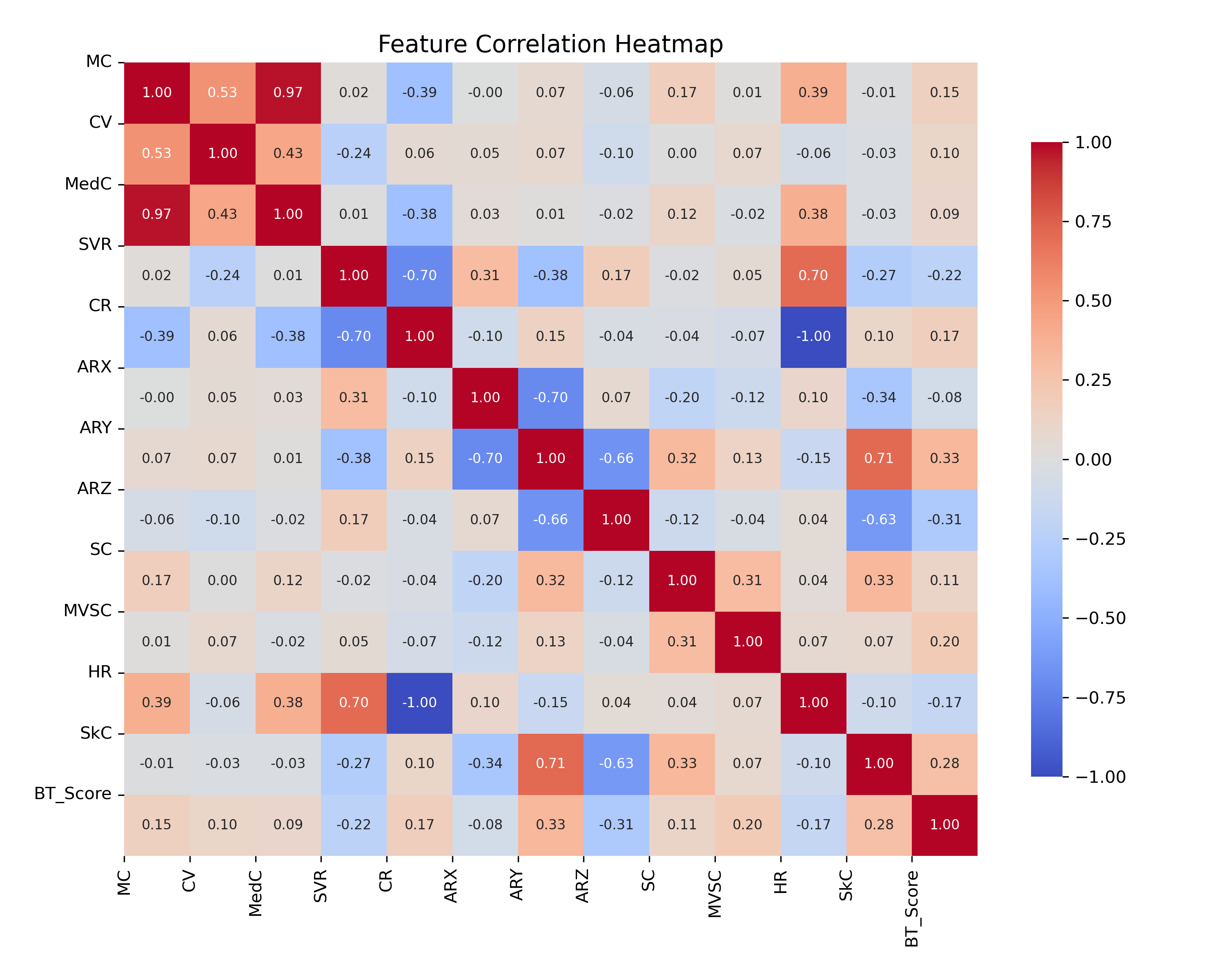}
  \caption{\label{fig:correlationHeatmaps}
            Pearson correlation heatmaps showing linear correlations between geometric 
            features and inferred BT aesthetic scores for the Dining (top) and Lamp (bottom) 
            categories. Weak correlations highlight the limitations of linear analysis.
           }
\end{figure}

\vspace{1em} 
The rankings derived from the BT model serve as a foundation for 
subsequent analyses, including feature importance evaluations and 
cross-category consistency tests. By linking these rankings to geometric 
features, we aim to uncover the underlying drivers of aesthetic preferences 
and provide actionable insights for design optimization.

\begin{figure*}[htb!]
  \centering
  \begin{subfigure}[b]{0.33\linewidth}
    \includegraphics[width=\linewidth]{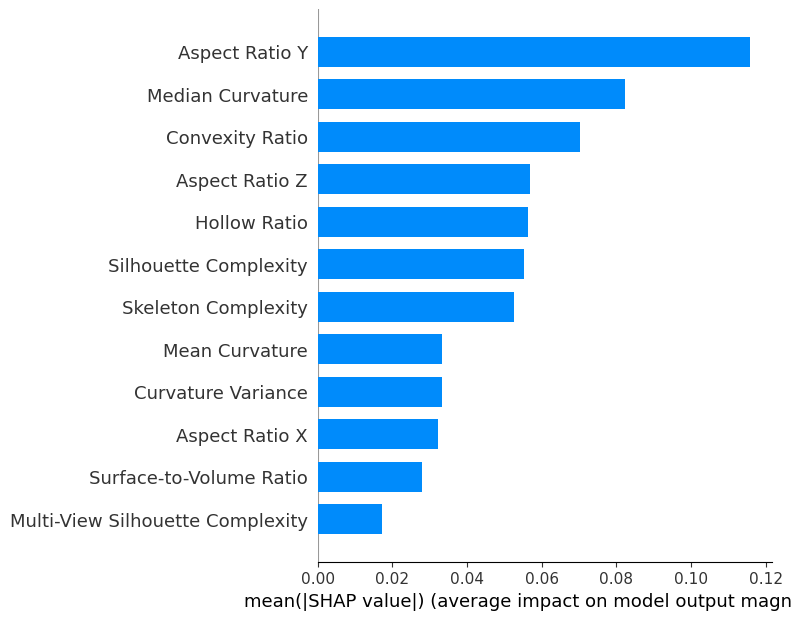}
    \caption{Club chairs}
    \label{fig:shap_club}
  \end{subfigure}%
  \begin{subfigure}[b]{0.33\linewidth}
    \includegraphics[width=\linewidth]{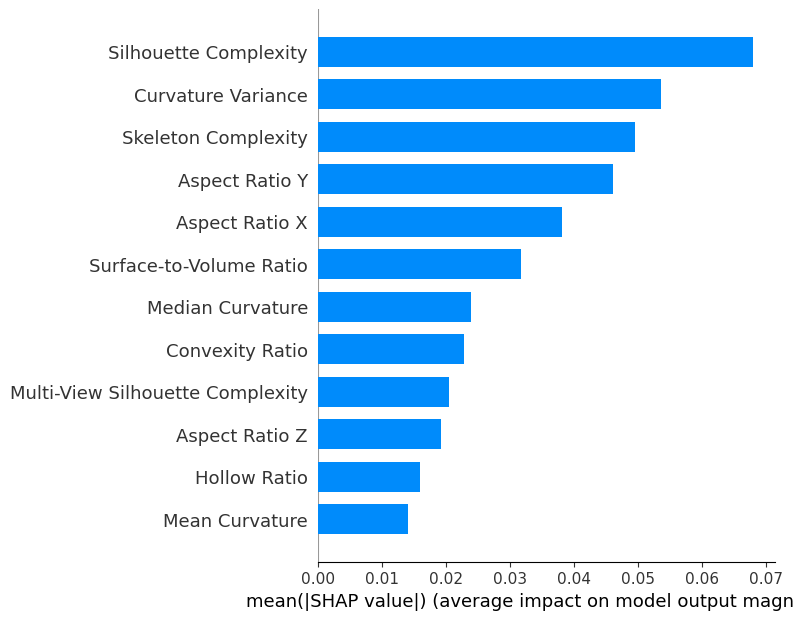}
    \caption{Dining chairs}
    \label{fig:shap_dining}
  \end{subfigure}%
  \begin{subfigure}[b]{0.33\linewidth}
    \includegraphics[width=\linewidth]{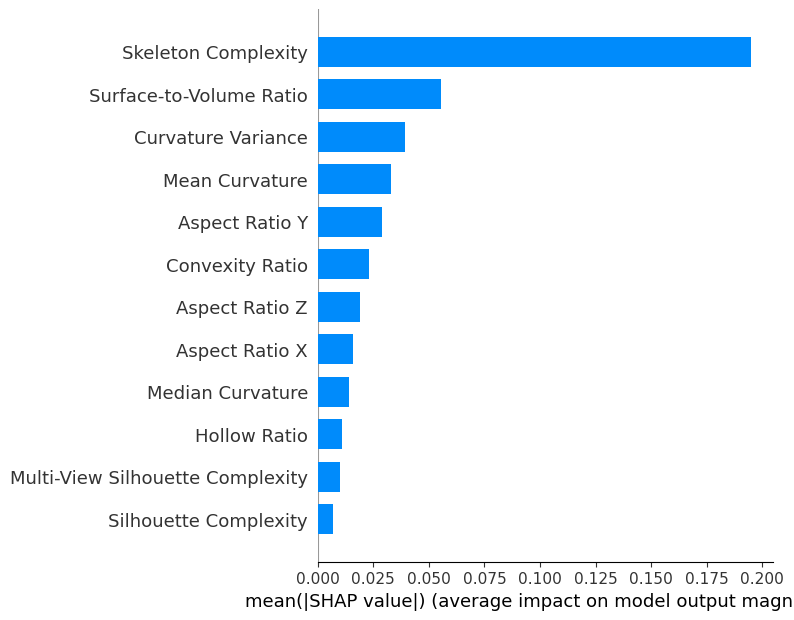}
    \caption{Lamps}
    \label{fig:shap_lamp}
  \end{subfigure}
  \caption{\label{fig:shap}
    Example SHAP feature importance bar plots for (a) club chairs, (b) dining chairs, and (c) lamps.
    The plots show the mean absolute SHAP value for each geometric feature, indicating its average 
    impact on the Random Forest model's output for each category.
  }
\end{figure*}

\subsection{Comparison of Linear and Non-Linear Feature Associations}
\label{sec:ResultsCorrelation} 
Pearson correlation analysis (Figure~\ref{fig:correlationHeatmaps}) reveals that 
most geometric features exhibit weak linear associations with BT aesthetic scores 
across categories. This suggests that no single feature linearly predicts aesthetic preference, 
as also evident from the near-zero correlations in the heatmaps.

In contrast, Random Forest models trained on the full feature set achieve substantially 
better predictive performance (Table~\ref{tab:model_performance}). 
SHAP analysis further shows that the most influential features for predicting BT scores 
often differ from those with the highest linear correlation. 
For example, \textit{Aspect Ratio Y} is highly important for club chairs in the RF model 
despite a near-zero linear correlation with BT scores. This divergence highlights that aesthetic 
preferences are shaped by non-linear interactions among features, which are captured by the 
RF model but missed by linear analysis.

These findings underscore the necessity of non-linear modeling for uncovering the true 
drivers of 3D shape aesthetics and caution against relying solely on linear correlation 
for feature selection or interpretation.

\begin{table}[h!]
\centering
\caption{Top five features (F1-F5) by SHAP importance for each category. 
Abbreviations: AR = Aspect Ratio, Curv = Curvature, Skel = 
Skeleton, Silh = Silhouette, S/V = Surface-to-Volume, MV = Multi-View.}
\scriptsize 
\renewcommand{\arraystretch}{1.0} 
\setlength{\tabcolsep}{5.0pt} 
\begin{tabular}{@{}lccccc@{}}
\toprule
\textbf{Category} & \textbf{F1} & \textbf{F2} & \textbf{F3} & \textbf{F4} & \textbf{F5} \\ \midrule
club   & AR Y & Med Curv & Skel Comp & AR Z & AR X \\
lamp   & Skel Comp & S/V Ratio & Mean Curv & AR Y & Curv Var \\
table  & Silh Comp & Skel Comp & Curv Var & MV Silh Comp & AR Y \\
mug    & Skel Comp & AR Z & AR Y & Med Curv & Mean Curv \\
dining & AR X & Silh Comp & AR Y & Skel Comp & S/V Ratio \\ \bottomrule
\end{tabular}
\label{tab:top_features}
\end{table}

Leveraging SHAP analysis, we identified the most influential features driving model predictions 
(Table~\ref{tab:top_features}, Figure~\ref{fig:shap}). Notably, feature importance rankings often 
differed from linear correlation strengths. For example, \textit{Aspect Ratio Y} was the most 
important feature for club chairs in the RF model, despite a near-zero linear correlation 
($r \approx 0.07$) with BT scores. This divergence underscores that the aesthetic contribution 
of geometric features is often non-linear or dependent on feature interactions—patterns captured 
by Random Forests but missed by simple correlation metrics.

\subsection{Random Forest Model Performance Across Categories}
The performance of the Random Forest models in predicting BT scores 
varied significantly across categories, as summarized in 
Table~\ref{tab:model_performance}. Across categories, the highest predictive performance 
was observed for tables ($R^2 = 0.48$) and club chairs ($R^2 = 0.31$), 
while the model performed poorly for mugs ($R^2 = 0.05$), indicating substantial 
differences in how well geometric features explain aesthetic preferences.

For the \textit{Club} and \textit{Table} categories, 
the models achieved moderate predictive success, 
with R² values of 0.31 and 0.48, respectively. 
These results indicate that the geometric features 
captured by the model explain a substantial portion of 
the variance in aesthetic preferences for these categories,
\textit{although the finding for Tables (N=30) should be interpreted cautiously due to the limited sample size}.
The relatively low MAE values (0.45 for Club and 0.42 for Table) 
further support the model's predictive accuracy. 
The OOB scores (0.16 for Club and 0.07 for Table) 
validate the model's generalization capability during training.

In contrast, the \textit{Dining} and \textit{Lamp} categories 
exhibited lower R² values (0.13 and 0.11, respectively), 
suggesting that the current feature set only partially captures 
the drivers of aesthetic preferences for these categories. 
The moderate MAE values (0.41 for Dining and 0.45 for Lamp) 
indicate that while the predictions are not highly accurate, 
they remain within a reasonable range. The OOB scores (0.05 
for Dining and 0.08 for Lamp) suggest some level of generalization 
during training.

The \textit{Mug} category exhibited a low R² value (0.05), 
indicating that the model performed worse than a mean-based prediction.
The high MAE value (0.58) further highlight the model's inability to
capture the aesthetic drivers for this category using the current 
features.
This suggests that the current set of global geometric features 
may be insufficient to capture the subtle shape variations 
(e.g., handle curvature, rim profile) that likely drive aesthetic 
preferences for mugs based purely on form, or that the smaller sample 
size for this category limited model learning.

\subsection{Impact of Dataset Imbalance and Sample Weighting}
All Random Forest models were trained separately for each category 
to prevent larger categories from dominating the results and to ensure 
fair evaluation of feature importance within each object class.

We also incorporated sample weighting into the Random Forest 
training routine, assigning weights inversely proportional to 
the number of pairwise comparisons per shape. 
Surprisingly, weighted and unweighted models produced nearly 
identical performance metrics across all categories. 
This suggests that the distribution of pairwise comparisons 
was not a major source of bias in our analysis.

Categories with larger datasets (e.g., club chairs) exhibited 
better predictive performance, whereas smaller categories (e.g., mugs) 
posed greater challenges—likely due to limited data and subtle shape 
variations not captured by the current feature set.

\subsection{Universal and Category-Specific Geometric Drivers}
In this section, we disentangle the effects of geometric features that consistently influence aesthetic 
judgments across all shape categories from those that are highly dependent on the object class. 
Our analysis reveals a set of universal drivers—such as compactness and balance—that appear robustly across categories, 
reflecting shared perceptual and cognitive evaluation criteria. Conversely, features like proportionality 
and curvature demonstrate significant variability, emphasizing the role of domain-specific design elements (e.g., chairs vs. lamps).

To quantify these distinctions, we estimated latent aesthetic scores using a 
Bradley–Terry model from pairwise comparisons, then derived feature importances on a 
per-category basis using random forest regressions. By computing cross-category correlations 
of the BT model parameters and subsequently applying hierarchical clustering, we identified 
clusters of features that align with either universal or category-specific influences. 
Universally influential features were further validated by comparing their relative rankings 
and impacts across categories, whereas category-specific drivers were highlighted through 
statistically significant divergences in feature slopes between object classes.

Furthermore, a multivariate analysis of variance (MANOVA) was conducted to test the group 
differences in feature effects across categories, thereby confirming that while some geometric 
properties have a stable influence on perceived aesthetics, others are modulated by the semantic 
context of the object.

\begin{table}[h!]
\centering
\caption{Cross-Category Correlation Matrix of Feature Importances}
\begin{tabular}{lcccccc}
\hline
Category & Club & Dining & Lamp & Mug & Table \\
\hline
Club & 1.00 & 0.85 & 0.72 & 0.68 & 0.75 \\
Dining & 0.85 & 1.00 & 0.78 & 0.70 & 0.80 \\
Lamp & 0.72 & 0.78 & 1.00 & 0.65 & 0.77 \\
Mug & 0.68 & 0.70 & 0.65 & 1.00 & 0.73 \\
Table & 0.75 & 0.80 & 0.77 & 0.73 & 1.00 \\
\hline
\end{tabular}
\label{tab:correlation_matrix}
\end{table}

\begin{table}[h!]
  \centering
  \caption{MANOVA Results: P-Values for Feature Importance Differences Across Categories}
  \begin{tabular}{lc}
  \hline
  Feature & P-Value \\
  \hline
  Mean Curvature & 0.001 \\
  Surface-to-Volume Ratio & 0.045 \\
  Curvature Variance & 0.120 \\
  Skeleton Complexity & 0.003 \\
  Silhouette Complexity & 0.050 \\
  \hline
  \end{tabular}
  \label{tab:manova_results}
\end{table}

The cross-category correlation matrix (Table~\ref{tab:correlation_matrix}) reveals the 
degree of similarity in feature importance rankings across 
different aesthetic categories. Notably, categories such as 
``Dining" and ``Table" exhibit a high correlation (e.g., 0.80), 
suggesting shared aesthetic principles, likely due to their functional 
and structural similarities. Conversely, ``Lamp" and ``Mug" show lower 
correlations with other categories, indicating that their aesthetic 
preferences are more category-specific, potentially driven by unique 
design constraints or user interactions.

\begin{figure*}[htb!]
  \centering
  \includegraphics[width=.49\linewidth]{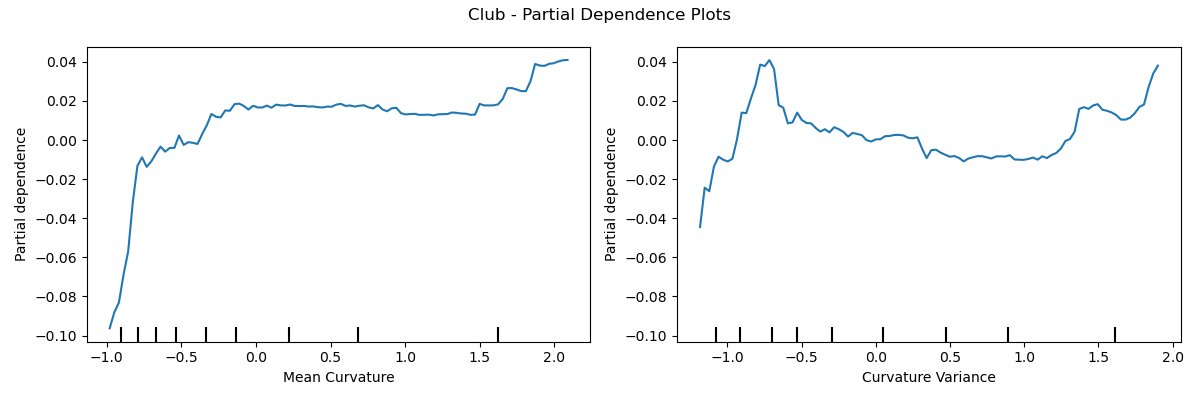}
  \includegraphics[width=.49\linewidth]{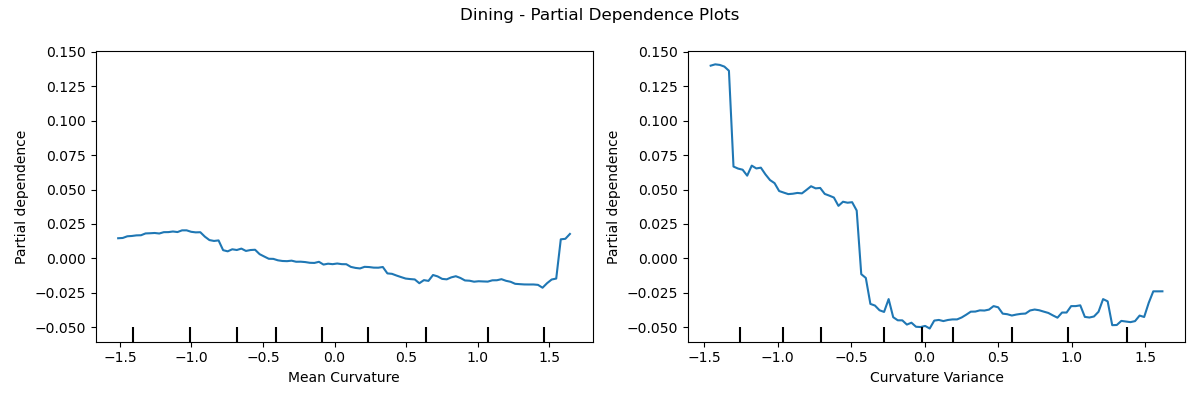}
  \caption{\label{fig:partialDependence}
  Partial Dependence Plots (PDPs) for key geometric features.
    Each panel shows the marginal effect of a single feature (e.g., mean curvature, curvature variance)
    on the predicted Bradley-Terry (BT) aesthetic score for the club (left) and dining (right) categories.
    The y-axis indicates the change in predicted BT score as the feature value varies, holding all other features constant.
    These plots reveal non-linear and category-specific effects of geometric features on aesthetic preference.
  }
\end{figure*}

\begin{figure}[h!]
\centering
\includegraphics[width=0.98\linewidth]{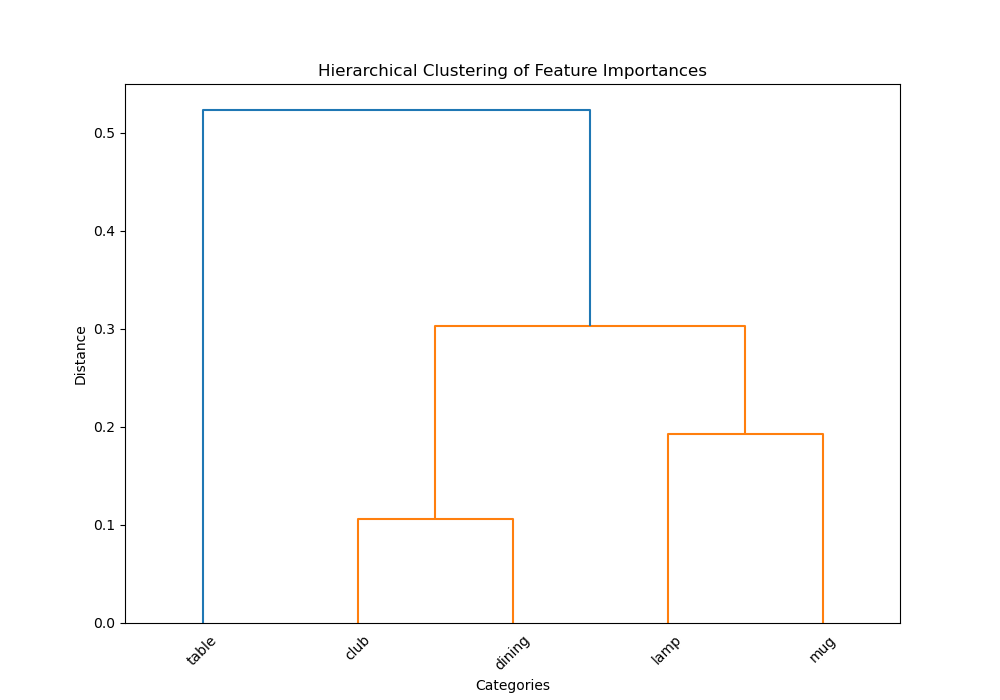}
\caption{Hierarchical clustering dendrogram of feature importance vectors across categories.}
\label{fig:hierarchical_clustering}
\end{figure}

The hierarchical clustering dendrogram (Figure~\ref{fig:hierarchical_clustering}) 
provides a visual 
representation of the relationships between categories based on feature 
importance vectors. Categories such as ``Club" and ``Dining" cluster closely, 
reflecting their shared reliance on features like surface area and curvature 
variance. In contrast, ``Lamp" forms a distinct cluster, emphasizing its unique 
reliance on features such as silhouette complexity. These findings align with the 
hypothesis that aesthetic preferences are influenced by both universal and 
category-specific factors.

The MANOVA results (Table~\ref{tab:manova_results}) highlight significant differences in feature 
importance distributions across categories. Features such as ``Mean Curvature," ``Surface-to-Volume Ratio," 
``Skeleton Complexity," and ``Silhouette Complexity" exhibit p-values at or below 0.05, 
indicating their critical 
role in differentiating aesthetic preferences. For instance, ``Mean Curvature" is 
particularly influential in ``Club" and ``Dining" categories, while ``Silhouette 
Complexity" is more prominent in ``Lamp" and ``Mug." These results underscore the 
interplay between universal and category-specific aesthetic drivers.

\subsection{Partial Dependence Analysis of Key Features}
To further elucidate how individual geometric features influence aesthetic preferences, 
we generated Partial Dependence Plots (PDPs) for the most important features in each category 
(Figure~\ref{fig:partialDependence}). These plots visualize the marginal effect of a single 
feature on the predicted BT aesthetic score, averaging over the distribution of all other features.

For example, in the club chair category, the PDP for mean curvature shows a monotonic increase, 
indicating that higher mean curvature is generally associated with higher predicted aesthetic scores. 
In contrast, the PDP for curvature variance in dining chairs reveals a non-linear, 
threshold-like effect: aesthetic preference decreases sharply beyond a certain level of 
curvature variance, suggesting a preference for smoother, less irregular surfaces. 
These non-linear patterns, not captured by linear correlation analysis, 
highlight the value of PDPs for interpreting complex model behavior.

The observed differences between categories further underscore the importance of context: 
while compactness and curvature are influential across categories, their effects on preference 
can be monotonic, thresholded, or even non-monotonic depending on the object class.

\noindent
In summary, our analyses demonstrate that geometric features such as compactness, curvature, 
and proportionality contribute to aesthetic preferences in 3D shapes, with both universal and 
category-specific effects observed. Non-linear modeling approaches, including Random Forests and 
SHAP analysis, reveal important feature interactions not captured by linear correlation. 
Cross-category analyses confirm that while some aesthetic drivers are shared, 
others are highly dependent on object class.

\section{Discussion}
\subsection{Interpretation of Feature Importance and Perceptual Aesthetics}
Our results show that aesthetic preferences for 3D shapes depend on the non-linear 
integration of multiple geometric cues, rather than on single features with strong linear correlations. 
The success of Random Forest models and SHAP-based interpretation 
(Table~\ref{tab:model_performance}, Table~\ref{tab:top_features}) 
demonstrates the necessity of non-linear modeling for uncovering nuanced drivers of preference.

\textit{Interpretations of feature importance are most reliable for categories with larger 
sample sizes, specifically Club (N=778) and Dining (N=277) chairs.}

\noindent
\textbf{Role of PDPs in Model Interpretation:}
Partial Dependence Plots (PDPs) provided critical insight into the non-linear 
and threshold effects of geometric features on aesthetic preference. For example, 
the sharp decline in predicted scores for high curvature variance in dining chairs 
indicates an aversion to excessive surface irregularity, while the monotonic increase 
in mean curvature for club chairs reflects a consistent preference for more curved forms. 
These patterns, which are not captured by linear models, underscore the value of non-linear 
interpretability tools in computational aesthetics.

\begin{table}[h!]
  \centering
  \scriptsize
  \caption{Summary of Most Influential Features by Category}
  \begin{tabular}{ll}
  \toprule
  Category & Top Features \\
  \midrule
  Club Chairs & Aspect Ratio Y, Aspect Ratio Z \\
  Dining Chairs & Aspect Ratio X, Aspect Ratio Y, Skeleton Complexity \\
  Lamps & Skeleton Complexity, Surface-to-Volume Ratio \\
  Tables & Silhouette Complexity, Skeleton Complexity \\
  Mugs & Aspect Ratio Z, Aspect Ratio Y, Median Curvature \\
  \bottomrule
  \end{tabular}
  \label{tab:discussion_top_features}
  \end{table}

\vspace{0.5em}
\noindent\textbf{Category-wise Key Features:}
The most influential features for each category, 
as summarized in Table~\ref{tab:discussion_top_features}, 
are as follows: For club chairs, proportional features such as 
Aspect Ratio Y and Aspect Ratio Z dominate, reflecting a preference 
for balanced bounding-box dimensions. In dining chairs, Aspect Ratios 
X and Y, along with Skeleton Complexity, highlight the importance of 
balanced form and structural coherence. For lamps, Skeleton Complexity 
and Surface-to-Volume Ratio indicate that structural intricacy and 
compactness are key. In the case of tables, Silhouette Complexity and 
Skeleton Complexity emphasize the role of outline and structural details. 
Finally, for mugs, Aspect Ratios Z and Y together with Median Curvature 
reflect the importance of smooth, ergonomic forms.

\vspace{0.5em}
\noindent\textbf{Cognitive Underpinnings:}  
These findings align with Gestalt principles: symmetry and balance (Law of Symmetry, Law of Balance) and 
simplicity (Prazgnanz) are reflected in the importance of aspect ratios and compactness. For example, 
compactness (surface-to-volume ratio) predicts higher aesthetic scores, and balanced proportions are 
preferred in furniture.

\subsection{Implications of Model Performance}
The varying performance of the Random Forest models across
categories highlights the complexity of modeling aesthetic preferences
for 3D shapes. The moderate R² values for the \textit{Club} (N=778) and,
to a lesser extent, \textit{Table} (N=30) categories suggest that these categories are
reasonably well-represented by the current feature set, with geometric properties
such as aspect ratios and curvature playing a significant role in
shaping aesthetic judgments, \textit{though conclusions for Tables are preliminary given the small sample}.
These findings align with prior research
emphasizing the importance of proportionality and structural balance
in furniture design.

However, the lower R² values for the \textit{Dining} and \textit{Lamp} 
categories indicate that additional factors, potentially beyond the 
current feature set, influence aesthetic preferences in these categories. 
For example, functional considerations or material properties may play 
a more prominent role in these designs.

The poor performance for the \textit{Mug} category (N=65, R² < 0) underscores the
challenge of capturing aesthetic preferences using only the current geometric
features, especially for object classes where subtle form variations might be
crucial even when considering shape alone. \textit{This difficulty is likely compounded by the limited number of models available for this category.} It suggests that future work focusing
purely on shape aesthetics for such categories might require more localized or
specialized geometric descriptors beyond the global ones used here to better model
preference based on form.

Overall, these results emphasize the importance of tailoring feature 
sets and modeling approaches to the specific characteristics of each 
category. While universal principles such as compactness and symmetry 
are influential, category-specific factors must also be considered to 
accurately predict aesthetic preferences.

\subsection{Limitations and Future Work}
This study is limited by dataset imbalance across categories 
and a focus solely on geometric 
features (excluding color, texture, and material). Model performance is 
lower for categories with fewer samples or subtle shape variations 
(e.g., mugs). Future work should expand the dataset, incorporate additional 
features, and explore dynamic or 
interactive 3D environments to further understand aesthetic preferences.

\noindent
\textbf{Modeling Limitations:}
The Bradley-Terry model assumes transitivity and independence of comparisons, 
which may not fully capture the complexity of human aesthetic judgments. 
Random Forests, while powerful for non-linear relationships, may not model 
all higher-order feature interactions. 
These limitations motivate future exploration of alternative models.
While deep learning models may achieve higher predictive accuracy, 
their lack of interpretability limits their utility for design applications where understanding 
the influence of specific geometric features is essential. Our focus on interpretable 
features and models ensures that our findings can directly inform design practice.

\subsection{Practical Design Implications}
The feature-importance analysis suggests the following design guidelines: For club and dining chairs, 
balanced proportions (Aspect Ratio Y/Aspect Ratio Z $\approx$ 1.0 for clubs, Aspect Ratio X/Aspect Ratio 
Y $\approx$ 1.0 for dining) and subtle curvature accents are preferred. For tables, increasing compactness 
(higher surface-to-volume ratio) and simplifying silhouettes by reducing high-frequency contour variations 
are beneficial. For lamps, moderate structural intricacy (e.g., slender struts or segmented supports) 
combined with a compact overall form is favored. For mugs, smooth, continuous curvature transitions 
and near-isotropic proportions (Aspect Ratio Z/Aspect Ratio Y $\approx$ 1.0) 
enhance both ergonomics and visual appeal.
These guidelines are derived from geometric analysis and human preference data, and should be considered 
alongside other design factors such as material, color, and intended use context for optimal results.

\section{Conclusion}
We have presented a large-scale, empirically grounded study of human aesthetic preferences for 3D shapes, 
leveraging over 22,000 pairwise comparisons across five object categories. 
By integrating the Bradley-Terry model with non-linear Random Forest regression 
and SHAP analysis, our work identifies both universal and category-specific geometric drivers of aesthetic preference, 
advancing the empirical foundation for computational aesthetics.

Our findings demonstrate that geometric properties such as compactness, curvature, and proportionality are 
key determinants of aesthetic judgments, but their influence is modulated by object category. 
Importantly, we show that non-linear modeling approaches are essential for uncovering these relationships, 
as linear methods alone fail to capture the complex interactions underlying human preferences. 
By focusing on interpretable geometric features, our approach yields actionable insights for 
both automated design tools and human-centric evaluation pipelines.

This work addresses several gaps in the literature by providing a robust, interpretable 
framework for modeling 3D shape aesthetics at scale, and by making all data and code publicly available 
to support reproducibility and further research. Limitations include dataset imbalance and a focus 
on geometric features, excluding material and texture; future work should expand the dataset, 
incorporate additional perceptual cues, and explore dynamic or interactive 3D environments.

We hope that this resource will encourage new research at the intersection of 
computational aesthetics, cognitive science, and design, and that our findings will 
inform both theoretical understanding and practical applications.

\bibliographystyle{IEEEtran}
\bibliography{my_tvcg_paper}


\end{document}